\documentclass{article}
\newcommand{\bfr}{\begin{flushright}}
\newcommand{\efr}{\end{flushright}}
 
\begin{document}
\title{Superradiance from a Charged Dilaton Black Hole
}
\author{Kiyoshi Shiraishi\\
Akita Junior College, Shimokitade-Sakura, Akita-shi, \\Akita 010,
Japan
}
\date{Modern Physics Letters {\bf A7} (1992) pp. 3449--3454
}
\maketitle
\begin{abstract}
We study the behavior of the wave function of charged Klein-Gordon field
around a charge dilaton black hole. The rate of spontaneous charge
loss is estimated for large black hole case.
\end{abstract}

\bigskip

Charged black holes lose their charges by spontaneous emission of
charged particles. This phenomenon can be attributed to the pair
creation of charged particles in the strong electric field near the
horizon. The rate of the reduction of the black hole charge agrees with
the Schwinger's formula in QED.\cite{1}

We can understand the phenomenon by investigating the wave function of
the charged quanta near the horizon of the black hole. In the
scattering process by a charged and/or rotating black hole, the
intensity of the reflected wave becomes larger than that of the
incoming wave under a certain condition for the frequency. This fact
has been called superradiance \cite{2,3,4} with great surprises. Since
this corresponds to the induced emission, the process corresponding to
the spontaneous emission must exist. The quantization of the field
around the black hole leads to such a process of spontaneous emission
of the quanta of the field. For Reissner-Nordst\"om black holes, the
emission of charged particle reduces the charge of black holes; the
process of the loss of the charge of the black holes was closely
investigated in Ref.~\cite{3}.

In this paper, we study the superradiant modes of charged scalar field
around a charged dilaton black hole.\cite{5,6,7,8,9} The features of the
dilaton black hole have been studied recently. The study on
thermodynamical property of the black holes \cite{5,6,7,8} reveals that
there is a critical value for the dilaton coupling, above which the
thermodynamical description of black holes seems to break down.\cite{8}
Thus the quantum emission of charged particles is naturally expected to
be affected by the dilaton field.

The metric for a spherically symmetric charged dilaton black hole in
$(1+N)$-dimensional space time is written in the form \cite{5,6,7,8}
\begin{equation}
ds^2=-\Delta\sigma^{-2}dt^2+\sigma^{2/(N-2)}(\Delta^{-1}dr^2+
r^2d\Omega^2_{N-1})\,,
\label{eq1}
\end{equation}
where
\begin{equation}
\Delta(r)=\left(1-\left(\frac{r_+}{r}\right)^{N-2}\right)
\left(1-\left(\frac{r_-}{r}\right)^{N-2}\right)
\end{equation}
and
\begin{equation}
\sigma^2(r)=
\left(1-\left(\frac{r_-}{r}\right)^{N-2}\right)^{2a^2/(N-2+a^2)}\,.
\end{equation}

In these equations $r_+$ and $r_-$ are integration constants and 
they are connected to the mass $M$ and electric charge $Q$ of the black
hole through the relations
\begin{equation}
\frac{16\pi M}{A_{N-1}(N-1)}=(r_+)^{N-2}+
\frac{N-2-a^2}{N-2+a^2}(r_-)^{N-2}
\label{eq4}
\end{equation}
and
\begin{equation}
Q^2=\frac{(N-1)(N-2)^2}{2(N-2+a^2)}(r_-r_+)^{N-2}\,,
\label{eq5}
\end{equation}
where $A_{N-1}=2\pi^{N/2}/\Gamma(N/2)$. The horizon is located at 
$r=r_+$ in this metric.

The configurations of the dilaton $\phi$ and electric fields are given
by
\begin{equation}
e^{4a\phi/(N-1)}=\sigma^2(r)
\end{equation}
and
\begin{equation}
F=\frac{Q}{r^{N-1}} dt\wedge dr\,.
\end{equation}

The scalar charge $\Sigma$ of the black hole is given by \cite{2}:
\begin{equation}
\Sigma^2=\frac{a^2(N-1)(N-2)^2}{2(N-2+a^2)^{2}} (r_-)^{2(N-2)}\,.
\end{equation}
Here $\Sigma$ is defined by the asymptotic behavior of the dilaton field
\begin{equation}
\phi\sim\left(\frac{N-1}{2}\right)^{1/2}\frac{\Sigma}{(N-2) r^{N-2}}\,.
\end{equation}

We consider an extra scalar field $\psi$ with mass $m$ and electric
charge $e$ in the metric (\ref{eq1}). If we take the coupling to the
dilaton field into consideration, the action for the scalar field $\psi$
is expressed as
\begin{equation}
S_s=\int d^{N+1}x
\sqrt{-g}[e^{-4ab\phi/(N-1)}|(\nabla_\mu+ieA_\mu)\psi|^2
+e^{-4ac\phi/(N-1)}m^2|\psi|^2]\, ,
\end{equation}
where $b$ and $c$ are parameters which describe the strength to the
coupling to the dilaton.

If we think of particular underlying theories, the parameters $a$, $b$
and $c$ are known to take special values. In the string theory, it is
well known that $a^2$ equals to one.
For the matter field, we find a relation that $b-c=1$, provided that
the kinetic and mass terms of the matter field have the same dilaton
coupling in the ``string'' metric.\cite{6}

In the context of Kaluza-Klein theory, the matter field in $(N+1)$
dimensions would come from a neutral scalar field in $(N+2)$
dimensions. We then have an infinite number of species of charged
scalar fields. In this case, the parameters $a$, $b$ and $c$ are given
by $N^{1/2}$, $0$ and $-1$, respectively. Note that $b-c$ equals to one.

We use the following form, which realizes the separation of variables,
\begin{equation}
\psi=\frac{u_l(r)}{r^{(N-1)/2}}S_{(l)}(\{\theta\})\,e^{-i\omega
t}\,,
\end{equation}
where  $S_{(l)}$ is the harmonic function  on  $S^{N-1}$, which 
is the eigenfunction of the Laplacian on $S^{N-1}$ with unit radius,
such that  $\Delta^{(N-1)}S_{(l)}=-l(l+N-2)S_{(l)}$.

From the equation of motion, we obtain the radial wave equation:
\begin{eqnarray}
& &\frac{d^2u_l}{dy^2}+\left\{\left(\omega-\frac{eQ}{(N-2)r^{N-2}}
\right)^2\sigma^{-4b+2(N-1)/(N-2)}-m^2\Delta\sigma^{-2(b+c)+2/(N-2)}\right.\nonumber
\\ &
&\quad-\frac{N-1}{2}\left(\frac{d}{dr}\ln\frac{\Delta}{\sigma^{2b}}\right)  
\frac{\Delta^2\sigma^{-4b}}{r^2}
\nonumber
\\ & &\quad\left.-\left(l(l+N-2)+\frac{(N-1)(N-2)\Delta}{4}\right)
\frac{\Delta\sigma^{-4b}}{r^2}\right\}u_l=0\,,
\label{eq12}
\end{eqnarray}
where the new coordinate $y$ is defined by
\begin{equation}
\frac{dy}{dr}=\frac{\sigma^{2b}}{\Delta}\,,
\end{equation}
where $y$ varies between $-\infty$ and $\infty$, while $r$ varies
between
$r_+$ and $\infty$.  Note that our coordinate $y$ is different from the
coordinate $r_*$ adopted in Ref.~\cite{8}.

Let  us consider the scattering of classical wave by the black hole.
The solution for (\ref{eq12}) has an asymptotic form at infinity
($y\rightarrow\infty\,(r\rightarrow\infty)$):
\begin{equation}
u\sim e^{-iky}+A\, e^{iky}\,,
\end{equation}
where $k=(\omega^2-m^2)^{1/2}$.

The behavior of $u$ near the horizon
($y\rightarrow-\infty\,(r\rightarrow r_+)$) is determined, on the
other hand, taking the boundary condition that there is only incoming
wave at the horizon. That is 
\begin{equation}
u\sim B \exp\{-i(\omega-e\Phi_+)\sigma_+^{-2b+(N-1)/(N-2)}y\}\, ,
\end{equation}
where $\Phi_+=Q/\{(N - 2)r^{N-2}\}$ and $\sigma_+=\sigma(r_+)$. (Note
that $\Delta(r_+)=0$.)

Using the constancy of the Wronskian, we can show
\begin{equation}
1-|A|^2=|B|^2(\omega-e\Phi_+)\sigma_+^{-2b+(N-1)/(N-2)}/k\, . 
\end{equation}
Therfore we find that the reflection coefficient $R=|A|^2$ becomes
larger than one, provided that $m<\omega<e\Phi_+$. In other words, the
amplication of wave occurs. This
phenomenon is the well known superradiance. We find that the condition
for the superradiance does not change its form regardless of the
dilaton coupling.

The quantisation of scalar field reveals the process of spontaneous emission of
charged particles. The rate of the loss of charge can be calculated using the mode
summation over the superradiant modes.

Now we consider the large black hole case, $mr_+\gg 1$. This case
corresponds to the situation that the strength of the electric field
varies gradually from place to place. We can utilize the WKB
approximation method in this case. The wave equation can be simplified
as
\begin{equation}
\frac{d^2u}{dy^2}+Wu=0\, ,
\end{equation}
with
\begin{equation}
W=\left(\omega-\frac{eQ}{(N-2)r^{N-2}}\right)^2 
\sigma^{-4b+2(N-1)/(N-2)}-m^2\Delta \sigma^{-2(b+c)+2/(N-2)}\,.
\end{equation}
Then the rate of the spontaneous emission is proportional to
\begin{equation}
A(\omega)=\exp\left(-2\int_\alpha^\beta
(|W|)^{1/2}dy\right)\,,
\end{equation}
where $\alpha$ and $\beta$ are defined by $W(\alpha)=W(\beta)=0$.
If we add the condition $b-c=1$, we find that the integration takes
the simple form
\begin{eqnarray}
& &\int_\alpha^\beta(|W|)^{1/2}dy=\frac{k}{N-2}\nonumber \\
& &\quad\times\int_{\alpha^{N-2}}^{\beta^{N-2}}
\frac{x^{(1-a^2)/(N-2+a^2)}(x-\alpha^{N-2})^{1/2}
(\beta^{N-2}-x)^{1/2}}{(x-r_+^{N-2})
(x-r_-^{N-2})^{\{(N-2)^2-a^2\}/(N-2)(N-2+a^2)}}dx\,.
\label{eq20}
\end{eqnarray}
The case with $a=0$, i.e., the Reissner-Nordstr\"om case is also
included here.

Although the integral (\ref{eq20}) cannot be expressed as a combination
of elementary functions generally, the explicit results can be obtained
in some restricted cases.

For $a=0$ and  $N=3$, one can show
\begin{equation}
\int_\alpha^\beta(|W|)^{1/2}dy=\frac{\pi
m^2\{eQ-(\omega-k)M\}}{k(\omega+k)}\,,\quad
\mbox{for}~a=0~\mbox{and}~N=3\, .
\end{equation}

On the other hand, for $a=1$ and $N=3$, one can show
\begin{eqnarray}
& &\int_\alpha^\beta(|W|)^{1/2}dy=\frac{\pi
m^2\{eQ-(\omega-k)M-(\omega+k)Q^2/(2M)\}}{k(\omega+k)}\,,\nonumber
\\
& &
\qquad\mbox{for}~a=1~\mbox{and}~N=3\, .
\end{eqnarray}

Note that one must apply respective relations such as (\ref{eq4}) and
(\ref{eq5}) to each case.
Using the result, we find the rate of the charge loss for $a=0$ and
$N=3$
\begin{eqnarray}
\frac{dQ}{dt}&=&-e\int_m^{e\Phi_+}S_{(l)}A(\omega)d\omega
\\
&\sim&-\frac{e^4Q^3}{r_+}\exp\left(-\frac{\pi m^2r_+^2}{eQ}\right)
\{1+O(eQ/m^2r_+^2)\}\, , \nonumber \\
& &\mbox{for}~a=0~\mbox{and}~N=3\,,
\end{eqnarray}
where we assume $m\ll e\Phi_+$.

Similarly, we find for $a=1$ and $N=3$
\begin{eqnarray}
\frac{dQ}{dt}&\sim&-\frac{e^4Q^3}{r_+}\exp\left(-\pi
m^2\left(\frac{r_+^2}{eQ}-\frac{2Q}{e}\right)\right)
\{1+O(eQ/m^2r_+^2)\}\, , \nonumber \\
& &\mbox{for}~a=1~\mbox{and}~N=3\,,
\end{eqnarray}
where we assume $Q^2 < 2M^2$. For the case with nearly extreme charge,
$Q^2=2M^2$, the WKB evaluation is no longer appropriate for this case.

By examining the integrand of (\ref{eq20}), we find that the value of
integration de creases rapidly as the charge of the black hole
approaches the extremal value, if $a^2\ge(N-2)^2$. This is due to
the power of $(x-r_-^{N-2})$ in the denominator of the integrand. Thus
the WKB method becomes inapplicable in such cases.

For other general cases, we give an approximate result here:
\begin{equation}
\frac{dQ}{dt}\sim-e(e\Phi_+)^Nr_+^{N-1}\exp
\left(-\pi\frac{E_c}{E_+}\right)
\{1+O(E_+/E_c)\}\, ,
\end{equation}
where $E_+=Q/r_+^{N-1}$ and $E_c=e/m^2$. Here we assume $r_+\gg r_-$.

In this brief report, we have derived the superradiant condition for
the frequency of wave function around a charged dilaton black hole and
have estimated the rate of charge loss of the dilaton black holes.

We have found that the condition for superradiant modes remains
unchanged, i.e., $\omega<e\Phi_+$, even if arbitrary dilaton couplings
are present.

For the large dilaton coupling, $a\ge(N-2)$, we have found that the
WKB method to estimate the rate of spontaneous emission cannot be
applied in the case with nearly extremal electric charge. The string
theory in four dimensions and Kaluza-Klein theories fall under this
category.

Therefore we can naturally guess that the nonzero dilaton coupling enhances
the spontaneous emission of charged particles, especially in nearly extreme cases.

To investigate the critical stage in the reduction of charge of the dilaton black
hole, we need a numerical computation in any case. We should further study the
quantum process near the horizon of the dilaton black hole, especially in nearly
extremal case.


\end{document}